\documentclass[journal]{IEEEtran}
\usepackage{cite}
\usepackage[flushleft]{threeparttable}
\usepackage{booktabs}
\usepackage[caption=false]{subfig}
\usepackage{comment}
\usepackage{amsmath,amssymb,multirow}
\usepackage{graphicx}
\usepackage{algorithm}
\usepackage{algpseudocode}
\usepackage{xcolor}
\usepackage{amssymb}
\usepackage{pifont}
\usepackage{xspace}
\usepackage{listings}
\usepackage{xcolor}

\def\BibTeX{{\rm B\kern-.05em{\sc i\kern-.025em b}\kern-.08em
    T\kern-.1667em\lower.7ex\hbox{E}\kern-.125emX}}

\definecolor{codegreen}{rgb}{0,0.6,0}
\definecolor{codegray}{rgb}{0.5,0.5,0.5}
\definecolor{codepurple}{rgb}{0.58,0,0.82}
\definecolor{backcolour}{rgb}{0.95,0.95,0.92}

\lstdefinestyle{mystyle}{
    backgroundcolor=\color{backcolour},   
    commentstyle=\color{codegreen},
    keywordstyle=\color{magenta},
    numberstyle=\tiny\color{codegray},
    stringstyle=\color{codepurple},
    basicstyle=\ttfamily\footnotesize,
    breakatwhitespace=false,         
    breaklines=true,                 
    captionpos=b,                    
    keepspaces=true,                 
    numbers=left,                    
    numbersep=5pt,                  
    showspaces=false,                
    showstringspaces=false,
    showtabs=false,                  
    tabsize=2
}

\newsavebox{\LstBox}

\usepackage{amsmath,amsfonts,bm}

\newcommand{\beqn}{\begin{eqnarray}}
\newcommand{\eeqn}{\end{eqnarray}}
\newcommand{\beqnx}{\begin{eqnarray*}}
\newcommand{\eeqnx}{\end{eqnarray*}}

\def\va{{\bm{a}}}
\def\vb{{\bm{b}}}
\def\vc{{\bm{c}}}

\def\ve{{\bm{e}}}

\def\vk{{\bm{k}}}

\def\vn{{\bm{n}}}

\def\vq{{\bm{q}}}

\def\vs{{\bm{s}}}

\def\vz{{\bm{z}}}

\def\mA{{\bm{A}}}

\def\mC{{\bm{C}}}

\def\mE{{\bm{E}}}

\def\mK{{\bm{K}}}

\def\mQ{{\bm{Q}}}
\def\mR{{\bm{R}}}
\def\mS{{\bm{S}}}

\def\mV{{\bm{V}}}
\def\mW{{\bm{W}}}
\def\mX{{\bm{X}}}
\def\mY{{\bm{Y}}}

\DeclareMathAlphabet{\mathsfit}{\encodingdefault}{\sfdefault}{m}{sl}
\SetMathAlphabet{\mathsfit}{bold}{\encodingdefault}{\sfdefault}{bx}{n}

\def\sB{{\mathbb{B}}}

\def\sL{{\mathbb{L}}}

\def\sP{{\mathbb{P}}}

\def\sR{{\mathbb{R}}}
\def\sS{{\mathbb{S}}}

\def\sU{{\mathbb{U}}}

\def\softmax{{\mathrm{softmax}}}

\def\round{{\mathrm{round}}}
\def\fsq{{\mathrm{FSQ}}}
\def\normalize{{\mathrm{normalize}}}

\def\floor#1{\lfloor #1 \rfloor}
\def\1{\bm{1}}

\def\neer{$\mbox{NEER}$\xspace}
\title{Optimizing Contextual Speech Recognition Using Vector Quantization for Efficient Retrieval}

\author{\IEEEauthorblockN{Nikolaos Flemotomos$^\dagger$, Roger Hsiao$^\dagger$, Pawel Swietojanski$^\dagger$, Takaaki Hori, Dogan Can, Xiaodan Zhuang} \\
\IEEEauthorblockA{
\textit{Apple}\\
Cupertino, USA \\ 
\texttt{\{nflemotomos, rhsiao, pswietojanski\}@apple.com}}
}

\begin{document}
\bstctlcite{IEEEexample:BSTcontrol}

\maketitle

\def\thefootnote{$\dagger$}
\footnotetext{equal contributions; alphabetical order}
\def\thefootnote{\arabic{footnote}}

\begin{abstract}
Neural contextual biasing allows speech recognition models to leverage contextually relevant information, leading to improved transcription accuracy. However, the biasing mechanism is typically based on a cross-attention module between the audio and a catalogue of biasing entries, which means computational complexity can pose severe practical limitations on the size of the biasing catalogue and consequently on accuracy improvements. This work proposes an approximation to cross-attention scoring based on vector quantization and enables compute- and memory-efficient use of large biasing catalogues. We propose to use this technique jointly with a retrieval based contextual biasing approach. First, we use an efficient quantized retrieval module to shortlist biasing entries by grounding them on audio. Then we use retrieved entries for biasing. Since the proposed approach is agnostic to the biasing method, we investigate using full cross-attention, LLM prompting, and a combination of the two. We show that retrieval based shortlisting allows the system to efficiently leverage biasing catalogues of several thousands of entries, resulting in up to 71\% relative error rate reduction in personal entity recognition. At the same time, the proposed approximation algorithm reduces compute time by 20\% and memory usage by 85-95\%, for lists of up to one million entries, when compared to standard dot-product cross-attention. 
\end{abstract}
\begin{IEEEkeywords}
contextual biasing, vector quantization, speech recognition, finite scalar quantization, retrieval
\end{IEEEkeywords}

\section{Introduction}
\label{sect:intro}

\IEEEPARstart{E}{nd-to-end} deep learning models have revolutionized the field of automatic speech recognition (ASR), offering both simplicity and impressive performance~\cite{prabhavalkar2023end}. However, even large ASR models trained on vast amounts of data still struggle to recognize rare words---in particular names of entities~\cite{guo2019spelling, sainath2018no, alon2019contextual}. One way to address this problem is to allow models to leverage user-specific information, or context, during inference. Such context is often referred to as the user's biasing catalogue, and can include contact names, commonly used apps, media titles and creators, or relevant geo-locations. 

To that end, several approaches have been proposed in the literature, including shallow language model (LM) fusion~\cite{zhao2019shallow, le2021deep} and on-the-fly LM-based rescoring~\cite{hall2015composition, williams2018contextual}. Such methods, though, are often sub-optimal, because of the well-documented tendency of end-to-end models to learn a strong internal LM~\cite{mcdermott2021dr, meng2021internal}, and the fact that the contextual model is typically trained independently from the acoustic model (AM). Neural contextual biasing (NCB) offers an alternative paradigm where the biasing mechanism is part of the ASR model and jointly learned with the main ASR objective~\cite{pundak2018deep, chang2021context, sathyendra2022contextual, bruguier2019phoebe, jain2020contextual, sun2021tree, munkhdalai2022fast, meng2024text}.

NCB architectures usually rely on an additional context encoder followed by a fusion mechanism. The context encoder may be implemented as an LSTM~\cite{hochreiter1997long}, or more recently as a transformer~\cite{vaswani2017attention} module, and its role is to map a set of tokenized biasing phrases to a set of fixed-length continuous phrase embeddings that are integrated into ASR model predictions. This integration typically takes place in a latent space using cross-attention between audio and context encoders~\cite{pundak2018deep, jain2020contextual}. Another approach relies on utilizing large language models (LLMs) for contextualization. In this case, an LLM is used jointly with the ASR model and the biasing entries are packed into a prompt~\cite{minaee2024large} guiding LLM predictions~\cite{gong2024contextual,Wang2023_slm,chen2024salm,lei2024contextualization}.

Neural contextual biasing based on LLM prompting (thus on self-attention) or standard cross-attention fusion comes at the cost of increased compute and memory requirements, which means its applicability is limited to relatively small biasing catalogues in practical settings~\cite{Alexandridis2023,yang23o_interspeech,pundak2018deep,gong2024contextual}. On the one hand, LLMs have a fixed maximum context length, typically in the range of a few thousand tokens~\cite{minaee2024large}. Even when maximum context length is not a blocker, 
LLM inference cost can be prohibitive for long contexts~\cite{minaee2024large}, and LLM reasoning performance can degrade rapidly as context length increases~\cite{levy-etal-2024-task}. On the other hand, cross-attention, generally implemented as scaled dot-product~\cite{vaswani2017attention}, needs to be applied between all contextual embeddings (keys) and all acoustic encodings (queries), which can be restraining for large biasing catalogues. These limitations have led the research community to typically explore NCB settings where the biasing inventory is capped to a few thousand entries \cite{pundak2018deep,tong2023slot,chen2019joint,Kulshreshtha2023,sudo2024contextualized}.
\IEEEpubidadjcol

In this work we address the attention-driven computational limitations of NCB employing a quantization-based, two-stage approach. 
We discretize contextual embeddings using finite scalar quantization (FSQ), a variant of vector quantization~\cite{vq_2017} recently introduced in generative machine learning architectures~\cite{fsq_2024}. Utilizing certain properties of FSQ, we can approximate cross-attention in a memory- and runtime-efficient manner, and use this efficient implementation to accurately retrieve relevant biasing entries. These entries are then either (i)~integrated with acoustic encodings through a dot-product cross-attention biasing mechanism, or (ii)~used for LLM prompting in a delayed fusion setup~\cite{hori2025_df}.  

The main contributions of this paper are as follows: 
\begin{enumerate}
    \item We introduce vector quantization in the field of contextual speech recognition as a viable technique to discre\-tize biasing embeddings and approximate the compute-heavy cross-attention mechanism, achieving over 20\% speed boost and 85-95\% memory usage reduction.
    \item We use our proposed technique to efficiently retrieve a short list of entries from a large biasing catalogue. We show that this approach can offer up to 71\% relative error rate reduction for personal entity recognition.
    \item We pair our retrieval approach with different biasing implementations (full cross-attention, LLM prompting) and decoding configurations (auto-regressive, non auto-regressive), and demonstrate that our approach is flexible and agnostic to the specifics of the biasing mechanism or decoding algorithm.
    \item Our work opens up a path to scaling neural contextualization to relatively unexplored scenarios such as biasing towards large media catalogues, where the number of entries could be in the hundreds of thousands or millions.
\end{enumerate}

\section{Background and Related Work}
\label{sect:rel_work}

\subsection{Cross-attention based contextualization}
\noindent
Typical transformer-based NCB models employ a cross-attention fusion layer where the encoded biasing phrases are integrated with the acoustic encodings~\cite{jain2020contextual, munkhdalai2022fast}. Although traditionally applied at the output of the audio encoder, some recent works report improved accuracy by doing fusion via one or more intermediate audio encoder layers~\cite{dingliwal2023personalization,peng-etal-2024-owsm,huang2024contextual,Torres2024promptformer}. Cross-attention fusion has also been extended to architectures containing additional branches and losses to detect the biasing phrases~\cite{huang23d_interspeech,sudo2024contextualized}, and copying mechanisms able to directly copy the relevant biasing entries to the output transcription~\cite{jayanthi-etal-2023-retrieve,zhou-etal-2024-copyne}. Other efforts have focused on building more robust NCB systems by employing alternative sampling strategies to construct the biasing lists during training~\cite{jalal2023locality}, by applying hard negative mining~\cite{bleeker2023approximate}, or by injecting additional text-only information from the same distribution as the contextual domain~\cite{sainath2023improving,meng2024text}. 

\subsection{LLM based contextualization}
\noindent
LLMs are transformer-based models with billions of parameters that have yielded exciting results in general-purpose natural language understanding and generation~\cite{minaee2024large}. The speech community has been exploring ways to incorporate the power of LLMs into ASR decoding~\cite{rubenstein2023audiopalm,seide2024speech,yang2024promptasr,lakomkin2024end,chen2024its}, with shallow fusion~\cite{hu2023massively}, n-best rescoring~\cite{xu2022rescorebert}, and error correction re-decoding~\cite{yang2023generative} being some of the proposed approaches. Delayed fusion (DF) has also been recently introduced as an alternative~\cite{hori2025_df}, allowing one to use already trained LLMs and ASR models (potentially with very different vocabularies) to improve ASR quality without a large inference overhead.

LLM predictions can be further improved by \mbox{injecting} query-specific information via prompts~\cite{minaee2024large}. Generally \mbox{speaking}, during prompting, a compatible model gets conditioned on some number of prefix tokens, either at the attention decoder~\cite{pmlr-v202-radford23a} or encoder~\cite{peng-etal-2024-owsm,yang2024promptasr} level. Several works have recently tried to take advantage of the prompting mechanism of LLMs as an alternative contextualization approach. For instance,~\cite{chen2024salm} shows improved results transcribing technical talks when a small set of relevant keywords is included in the prompt, while~\cite{gong2024contextual} follows a similar approach, but reports a rapid degradation as the size of the biasing catalogue gets bigger. Prompting functionality has also been used in conjunction with retrieval mechanisms that operate at the phonetic~\cite{lei2024contextualization} or the acoustically encoded sentence level~\cite{Wang2023_slm}.

\subsection{Towards efficient contextualization}
\noindent
Attention is a ubiquitous technology across the machine learning community~\cite{niu2021review} and significant effort has been spent on developing efficient attention implementations. For instance, flash attention~\cite{dao2022flashattention} introduces substantial speed gains by optimizing the read and write operations between different levels of memory, while ring attention~\cite{liu2023ring} offers an elegant way to parallelize attention computation across multiple devices following block-wise processing. Such approaches, though, are hardware-oriented and do not address the fundamental memory requirements of attention computation.

In an effort to improve the computational efficiency of the biasing mechanism, some works dynamically limit the number of relevant phrases throughout decoding based on prefixes that appear in the partial hypotheses, applying bias-conditioning~\cite{pundak2018deep} or representing the biasing lists with prefix-trees~\cite{le2021deep,le21_interspeech,sun2021tree}. Other works have proposed to constrain NCB to be only triggered on a subset of relevant audio frames, something that also helps with the phenomenon of over-biasing~\cite{tong2023slot, Alexandridis2023}. Such approaches, despite limiting the number of time steps for which the dot-products need to be computed, still do not offer a scalable solution with respect to the length of the biasing list. 


One way to handle a large biasing catalogue in a flexible manner is to follow a two-stage, retrieval-based approach. In that line of work, the authors in~\cite{han2022improving,Munkhdalai2023nam,Wu2023DualNAM,wu2024context,meng2024text} apply a phrase-level attention to select a few candidate biasing phrases before applying a more expensive wordpiece-level attention~\cite{Wu2023DualNAM,wu2024context}. However, if $N$ denotes the number of biasing phrases and $k$ the number of wordpieces per phrase, this approach tries to find a trade-off between a latency of $O(N)$ and $O(kN)$, thus still struggles to scale well with the number of biasing entries, $N$.  
Another two-stage method is introduced in~\cite{yang23o_interspeech}, where the phone sequence of the streaming ASR output is used to filter out irrelevant entries of the biasing list. Phonemic information is also explored in~\cite{lei2024contextualization}, where a few candidate entities from a first pass are used to guide an LLM-based retrieval of phonologically similar entries and prompt the LLM during a second-pass decoding. A closely related work is dual attention~\cite{Fu2019dual_attention}, a technique recently applied to keyword spotting~\cite{sahai2023dual}. The idea relies on having two attention modules, a small and a large (full) one, and conditionally executing one of them depending on triggering signals. 

Retrieval approaches for ASR contextualization often operate in the linguistic~\cite{Bolaji2023_retr} or acoustic embedding space~\cite{wang2023_s2t_retrieval, Wang2023_slm}, where the embeddings act as queries to retrieve $k$ nearest neighbours (kNN)~\cite{Bolaji2023_retr} from a pre-built entity index, optionally using an auxiliary retrieval model~\cite{wang2023_s2t_retrieval, huang2024_retr}. Recently, a joint approach has been explored, where the retrieval step is part of the contextual ASR model~\cite{wu2024context,huang2024_retr}; in this case a form of dot-product score is still used to shortlist biasing entries, potentially employing an additional retrieval-oriented loss. The kNN and attention approaches are compared in~\cite{gourav2024multi}, for the case of multi-modal LLMs.
Index-based retrieval approaches allow for an efficient search, but introduce the cost of building and maintaining an up-to-date index, possibly aligned with auxiliary models that were used for indexing and/or retrieval. On the other side, joint model-level retrieval techniques are end-to-end learnable and self-contained, though at the higher cost of compute and memory requirements during inference. Our approach belongs to the latter category, though drastically limits the memory requirements during retrieval.

\subsection{Vector quantization}
\noindent
Vector quantization (VQ) is a data compression technique~\cite{vq_1984}, where a large set of points are represented by a finite smaller set of vectors, known as the codebook vectors. VQ codebooks can be automatically learned through gradient descent and used in conjunction with variational autoencoders to build robust generative models~\cite{vq_2017}. To that end, a VQ loss is added to the overall optimization objective, that essentially moves the codebook elements closer to the non-quantized points, based on their 
distances. Note that, instead of the learned quantized vectors, the original points can now be equivalently represented just by a single index in the learned codebook. 

A multitude of improvements to the original VQ formulation have been proposed in the literature~\cite{vqvae2_2019,rvq_2022,gvq_2023,peng2021generating,lee2022autoregressive}. The most notable ones in the field of audio processing are residual vector quantization (RVQ) and grouped VQ. RVQ~\cite{rvq_2022} employs a cascade of codebooks, where each one is responsible for the discretization of the quantization error, or residual, between the embedding and its quantized representation in the preceding layer, in a recursive manner. That way, the original points are represented by a set of indices (as many as the desired RVQ depth). Note that vanilla VQ would require exponentially larger codebooks to achieve the same quantization errors as RVQ. Grouped VQ~\cite{gvq_2023} also aims at decreasing the required number of codebooks and codebook vectors, by splitting the original embeddings into groups and employing separately trained VQs to quantize each group.

Despite the remarkable results achieved with VQ and the aforementioned variants, training such discrete latent representations remains challenging. 
Codebook collapse, the model learning to only use a very small subset of the available codebook vectors, is a well-documented problem. Several solutions have been proposed, including alternative distance metrics, carefully designed initialization and learning rate updates, embedding estimation through exponential moving averages, and auxiliary losses~\cite{roy2018theory,vqvae2_2019,lancucki2020robust,yu2022vectorquantized}.

In an effort to simplify the original VQ formulation, thus overcoming the associated challenges, the authors in~\cite{fsq_2024} introduced finite scalar quantization (FSQ) as an alternative. In FSQ, a $d$-dimensional vector (where $d$ is relatively small, with the original embedding typically being projected onto a lower dimension) is quantized by mapping each one of its $d$ values to an integer. To achieve that, the continuous values are first bounded through a non-linear function, and then rounded. Since there is a pre-defined set of levels \mbox{$\sL=[l_1, l_2, \cdots, l_{d}]$}, and the $i$-th dimension can only take one of $l_i$ different values (e.g., \mbox{$\{-1,0,1\}$} for $l_i=3$), the (implicit) codebook is of size $\prod_{i=1}^{d} l_i$. 
The idea of dropping explicitly trained codebooks is also explored in~\cite{lfq_2024}, where lookup-free quantization (LFQ) is introduced. LFQ operates similarly to a two-level FSQ, but training includes additional regularizations. 
\section{Proposed Architecture and Formulation}
\label{sect:formulation}

\subsection{Standard dense NCB}

\begin{figure*}[ht]
  \centering
  \subfloat[\label{fig:retrieval_ncb_vanilla}]{
  \includegraphics[trim={7.5cm 6.2cm 12.6cm 4.5cm},clip,width=.23\textwidth]{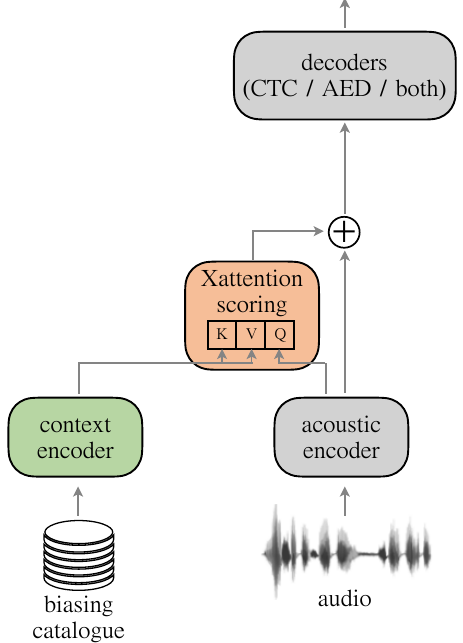}}
  \subfloat[\label{fig:retrieval_ncb_2pass}]{
  \includegraphics[trim={7.5cm 6.2cm 12.6cm 4.5cm},clip,width=.23\textwidth]{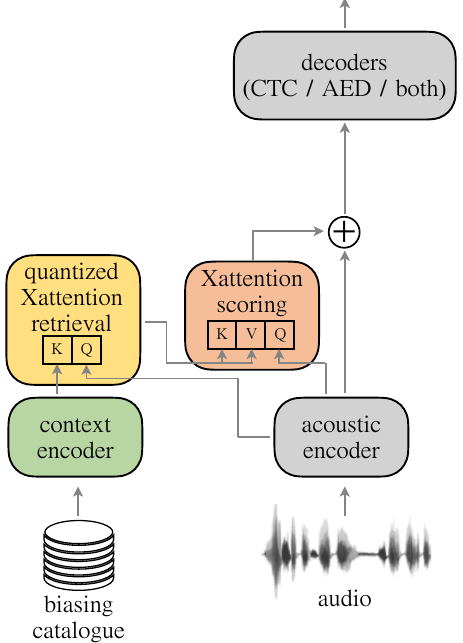}}
  \subfloat[\label{fig:retrieval_ncb_prompt}]{
  \includegraphics[trim={7.5cm 6.2cm 12.6cm 4.5cm},clip,width=.23\textwidth]{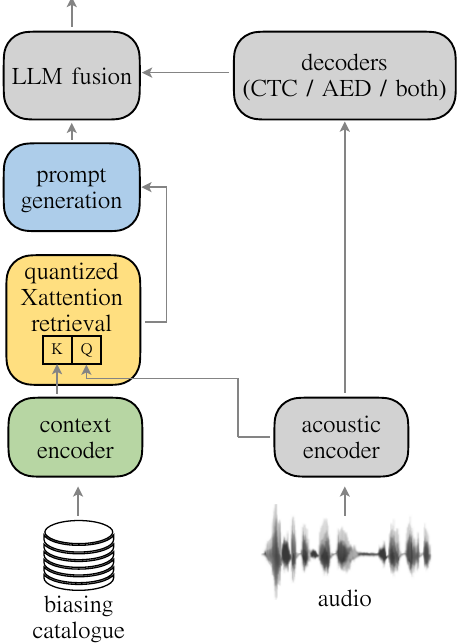}}
  \subfloat[\label{fig:retrieval_ncb_2passANDprompt}]{
  \includegraphics[trim={7.5cm 6.2cm 12.6cm 4.5cm},clip,width=.23\textwidth]{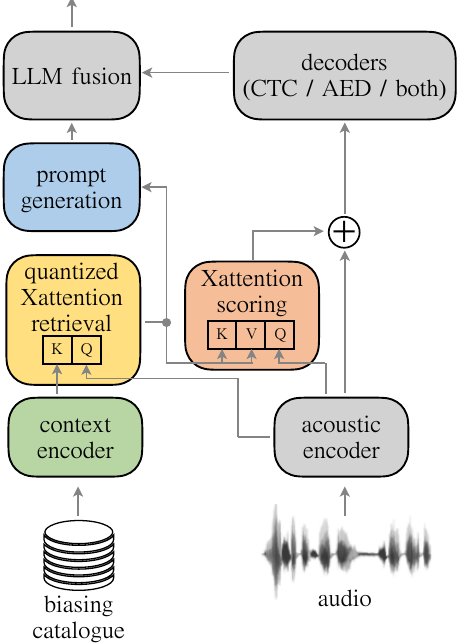}}
  \caption{Four considered biasing approaches: (a)~vanilla contextual biasing referred to as \emph{Dense NCB}; (b)~\emph{Retrieval NCB}, where a quantization module is used for large scale retrieval followed by TopK dense cross-attention processing; (c)~\emph{LLM Prompting}, where retrieved entries are packed into the prompt; (d)~combination of \emph{Retrieval NCB} and \emph{LLM Prompting}. Note that in an efficient implementation, the context encoder needs to be run exactly once for each biasing entry and the encodings get cached and reused across multiple queries.}
  \label{fig:retrieval_ncb}
  \vspace*{-.5cm}
\end{figure*}

\noindent
We build the proposed biasing system expanding on the CTC-AED model~\cite{KimHW17}, where the biasing mechanism is implemented using cross-attention 
between acoustic encodings and the biasing inventory. Formally, \mbox{$\mX\in\sR^{T\times D}$} denotes a sequence of acoustic embeddings, which are the output of a conformer-based acoustic encoder, and \mbox{$\mC\in\sR^{|\sB|\times D}$} denotes a list of contextual embeddings, as given by a transformer-based contextual encoder. Here $D$ is the output dimension of the encoders, $T$ is the length of the audio, and  $\sB$ is the list of biasing phrases, with $|\sB|$ denoting its length.\footnote{The acoustic features which are given as input to the acoustic encoder are often downsampled; $T$ represents the length of the downsampled sequence.} Note that the biasing list, $\sB$, always contains a special back-off token. The role of this special token is to help the model learn not to attend to a real biasing phrase whenever the utterance does not contain any contextual entity. Then, vanilla NCB computes
\begin{align}
    \{\mQ, \mK, \mV\} &= \{ \mX \mW_q, \mC \mW_k, \mC \mW_v\} \label{eqn:qkv_linearities}\\
    \mY &= \softmax\left(\alpha\;\mQ \mK^{\top}\right)\mV
    \label{eqn:cross_attention}
\end{align}
where $\mQ$, $\mK$, $\mV$ are known as queries, keys, and values respectively; $\mQ$ is computed by applying the linear transformation \mbox{$\mW_q\in\sR^{D\times D}$} to $\mX$; $\mK$ and $\mV$ are computed by applying the linear transformations \mbox{$\mW_k\in\sR^{D\times D}$} and \mbox{$\mW_v\in\sR^{D\times D}$}, respectively, to $\mC$; and \mbox{$\alpha=1/\sqrt D$} is a normalizing factor. The resulting biasing encodings, $\mY$, are added to the acoustic embeddings, $\mX$, and fed into CTC and/or AED decoders, similarly to~\cite{dingliwal2023personalization}. This process is referred to as \emph{Dense NCB} and is visually depicted in Fig.~\ref{fig:retrieval_ncb_vanilla}.

Although the above formulation 
stays relatively efficient for smaller biasing lists (say up to couple of thousand entries), the dot-product computation of Eq.~\eqref{eqn:cross_attention} could consume much compute and memory, so it could become a bottleneck, for larger catalogues. More precisely, we are computing \mbox{$T \cdot |\sB|$} $D$-dimensional dot-products. While both long audio and large inventory of biasing phrases could cause problems, the audio is often less of a concern since it can be chunked to enable online recognition, or the biasing can be applied only to a subset of audio frames using gating~\cite{Alexandridis2023}. The ability to handle a large inventory, $\sB$, however, cannot be easily dealt with. 

\subsection{Quantization and retrieval}

\noindent
Here we propose a two-stage attention module, where the first efficient implementation is based on vector quantization and is used for retrieving biasing entries grounded on the audio queries. The full attention is still used for biasing only with those relevant entries, as depicted in Fig.~\ref{fig:retrieval_ncb_2pass}. Alternatively, the retrieved biasing phrases can be used for prompting an LLM that contextualizes the final ASR predictions through delayed fusion, as shown in Fig.~\ref{fig:retrieval_ncb_prompt}. The two approaches can also be combined, as in Fig.~\ref{fig:retrieval_ncb_2passANDprompt}. 

Retrieval operates at the frame level, selecting the top entries from the dot-product between the queries and the keys. Such a frame-synchronous mode allows---depending on the target decoder---for both streaming and non-streaming applications. This differentiates our systems from approaches operating in the hypothesis space, retrieving additional contextual signals for second-pass decoding only~\cite{Zhou2024knn}, or systems assuming sentence-level representation is available for retrieval~\cite{Wang2023_slm}. Note that during the retrieval stage we do not need to compute the values, $\mV$, nor to calculate the softmax function in Eq.~\eqref{eqn:cross_attention}. While this further reduces the overall computational cost that would be otherwise required (in case of full cross-attention between audio queries and all the entries in the biasing list), the key novelty of our approach comes from the way we use vector quantization, and specifically FSQ, as detailed in the remainder of this section, and in Section~\ref{sect:vq_algo}.

In this work we apply FSQ to quantize the contextual embeddings of the NCB framework. In more detail, assume a contextual phrase with corresponding contextual embedding \mbox{$\vc \in \sR^D$}, as output from the NCB contextual encoder. Using FSQ with levels \mbox{$\sL=\left[ l_1, l_2, \ldots, l_{|\sL|}\right]$}, we get the quantized representation, \mbox{$\vz = \fsq(\vc)$}. However, depending on the exact FSQ setting, the resulting codebook capacity might be very limited. For instance, for \mbox{$\sL = [8,5,5,5]$}, the implicit codebook size is equal to \mbox{$\prod_il_i = 1000$}, whereas we want to represent hundreds of thousands or millions of potential biasing phrases. To increase the capacity of the quantizer, given a specific level configuration, $\sL$, we apply grouping and we quantize separately each one of the groups. The initial vector, $\vc$, is split into $G$ groups and we quantize each one of the \mbox{$\vc^{g} \in \sR^{(D/G)}$} sub-vectors. In our implementation, we use the same level configuration, $\sL$, for the quantizers applied to all the sub-groups (for all \mbox{$g\in\{1,2,\ldots,G\}$}). We get the final representation, $\vz$, as follows:
\begin{align}
    \tilde{\vc}^{g} &= \mA_{\text{in}}^g\vc^{g} + \vb_{\text{in}}^g
        = \left[ \tilde{c}_1^{g}, \tilde{c}_2^{g}, \ldots, \tilde{c}_{|\sL|}^{g} \right]&\forall g \label{eqn:fsq_in_projection}\\
    \ve^g &= \left[ \round [ f_1(\tilde{c}_1^g) ], \ldots, \round[ f_{|\sL|}(\tilde{c}_{|\sL|}^g) ] \right]&\forall g \label{eqn:fsq_bound_round}\\
    \vn^g &= \normalize\left( \ve^g \right) \label{eqn:fsq_normalize}&\forall g\\
    \vz^g &= \mA_{\text{out}}^g\vn^g + \vb_{\text{out}}^g&\forall g \label{eqn:fsq_out_projections}\\
    \vz &= \left[ {\vz^1}^{\top}; {\vz^2}^{\top}; \ldots; {\vz^G}^{\top} \right]^{\top} \label{eqn:fsq_regroup}
\end{align}
where \mbox{$[\mA_{\text{in}}^g ; \vb_{\text{in}}^g]$} is the input affine transformation that projects the \mbox{$\left(^D/_G\right)$-dimensional} vector $\vc^g$ onto an \mbox{$|\sL|$-dimensional} space; \mbox{$[\mA_{\text{out}}^g ; \vb_{\text{out}}^g]$} is the output affine transformation that projects the \mbox{$|\sL|$-dimensional} vector $\vn^g$ back onto the \mbox{$\left(^D/_G\right)$-dimensional} space; $\normalize(\cdot)$ is a function that maps the integer elements of $\ve^g$ to $[-1,1]$; $f_i(\cdot)$ is a bounding non-linear function such that each element $e_i^g$ of $\ve^g$ can take one of $l_i$ different integer values. We use the function proposed in~\cite{fsq_2024}, \mbox{$f_i(x) = \floor{^{l_i}/_2}\tanh{(x)}$}. The embedding, $\vc$, can be now represented by the $G$ indices to the implicit codebooks.

During training, we load the parameters of a pre-trained, non-quantized NCB model and we only train the additional FSQ parameters (i.e., the projections \mbox{$[\mA_{\text{in}}^g ; \vb_{\text{in}}^g]$} and \mbox{$[\mA_{\text{out}}^g ; \vb_{\text{out}}^g]\; \forall g$}), keeping the rest of the weights frozen.\footnote{Initial experimentation showed no improvements by training or fine-tuning all the network parameters jointly with the quantizer.} For back-propagation to work through the non-differentiable rounding operation, we copy gradients via straight-through estimation (STE)~\cite{bengio2013estimating}.\footnote{The rounding operation of a tensor, \texttt{x}, can be implemented in PyTorch with STE as \mbox{\texttt{x + (torch.round(x) - x).detach()}}.} 
Once the FSQ parameters have been trained, all the biasing phrases can be represented via the implicit FSQ codebooks, without the need to re-train any parameters based on the biasing catalogues at inference. 

As already mentioned, each element $e_i^g$ of $\ve^g$ can take one of $l_i$ different integer values for all the groups $g$; by extension, each element $n_i^g$ of $\vn^g$ can also take one of $l_i$ different values. Denoting as \mbox{$\sU = \{u_j\}_{j=1..|\sU|}$} the set of all possible values of \mbox{$n_i^g \; \forall i \forall g$}, we can easily see that, without loss of generality for the chosen bounding functions, the cardinality of the set is \mbox{$|\sU| \leq \sum_{i=1}^{|\sL|}l_i$}.\footnote{We use the inequality sign to allow for the (common) scenario where the same values are used across multiple levels.} With that in mind, we can represent each contextual embedding, $\vc$, by \mbox{$G\cdot|\sL|$} values from the set $\sU$. 
\section{VQ Based Algorithm for Cross Attention}
\label{sect:vq_algo}

\subsection{Efficient dot-product estimation}
\noindent
As explained in Section~\ref{sect:formulation}, we propose to apply FSQ to the contextual embeddings of the employed NCB architecture. This changes the structure of the dot-product appearing in Eq.~\eqref{eqn:cross_attention}. 
For simplicity, the formulation below assumes the number of groups is one, \mbox{$G=1$}, however, we found that grouping is important when it comes to the representational power of FSQ.\footnote{In our implementation, for \mbox{$G>1$}, we restricted $\mW_k$ to be a block-diagonal matrix with $G$ blocks, each one in $\sR^{(D/G)\times (D/G)}$.} Using a similar notation as in 
Eq.~\eqref{eqn:fsq_out_projections}---dropping the grouping indices---the dot-product between a query $\vq$ from $\mQ$ and a key $\vk$ from $\mK$, can be computed by
\begin{align}
\label{eqn:fsq_cross_attention1}
    \vq^{\top}\vk &\simeq \vq^{\top} \mW_k \vz \nonumber \\
        &= \vq^{\top} \left[ \mW_k (\mA_{\text{out}}\vn + \vb_{\text{out}}) \right] \nonumber \\
        &= \vq^{\top} \left[ \mA \vn + \vb \right] \nonumber \\
        &= \vq^{\top} \sum_{i=1}^{|\sL|} \va_i n_i + \vq^{\top}\vb  \nonumber \\
        &= \sum_{i=1}^{|\sL|} \vq^{\top}\va_i n_i + \vq^{\top}\vb
\end{align}
where $\vz$ is the quantized representation of the contextual embedding corresponding to $\vk$; $\mA$ and $\vb$ are equal to $\mW_k \mA_{\text{out}}$ and $\mW_k \vb_{\text{out}}$, respectively; $\va_i$ is the $i$-th column of $\mA$, i.e., \mbox{$\mA = [\va_1, \va_2, \ldots, \va_{|\sL|}]$}; $n_i$ is the $i$-th element of the \mbox{$|\sL|$-dimensional} FSQ vector representation $\vn$.  By examining the above equations closely, we notice:
\begin{enumerate}
    \item $\mA\in\mathbb{R}^{D\times|\sL|}$ and $|\sL|$ is relatively small,
    \item $n_i$ can take $l_i$ different values from $\sU$, 
    \item $\vq^{\top}\vb$ is not needed for retrieval. 
\end{enumerate}

Properties~1 and~2 have a big impact on computation and memory consumption because we no longer compute the dot-product matrix $\mQ\mK^{\top}$ between the queries and a big inventory of keys as in Eq.~\eqref{eqn:cross_attention}. Remember that $\mQ\mK^{\top}$ is a matrix in $\sR^{T\times|\sB|}$ and its estimation requires computing \mbox{$T \cdot |\sB|$} \mbox{$D$-dimensional} dot-products between $T$ different queries ($\vq$) and $|\sB|$ different keys ($\vk$). Instead of those $|\sB|$ different vectors, due to property~1, we now only have $|\sL|$ vectors (\mbox{$\{\va_i\}_{i={1..|\sL|}}$} in the notation of Eq.~\eqref{eqn:fsq_cross_attention1}), where $|\sL|$ is often in the range \mbox{4 -- 6}. Property~2 means there is limited variation for the values of $n_i$, since \mbox{$|\sU| \leq \sum_{i=1}^{|\sL|} l_i$}.\footnote{For the settings used in our experiments, \mbox{$\sL=[8,5,5,5]$} or \mbox{$\sL=[7,5,5,5,5]$}, so that \mbox{$|\sU|\leq 27$}.}  Combining properties~1 and~2 allows us to efficiently compute dot-products between the queries in $\mQ$ and all possible vectors from $\mA \vn$. Since we have $|\sL|$ columns in $\mA$ and each element in $\vn$ can take one of $|\sU|$ different values, we get at most \mbox{$|\sL| \cdot |\sU|$} \mbox{$D$-dimensional} vectors from $\mA \vn$. This would greatly reduce the memory consumption since \mbox{$|\sB| \gg |\sL| \cdot |\sU|$} for typical applications.

Property~3 implies we could ignore $\vq^\top \vb$ altogether since
for retrieval this term only adds a different constant to each time step and does not affect the TopK selection that takes place at the frame level. Therefore, we pre-compute the score matrix $\mS_{an} \in \mathbb{R}^{D \times (|\sL|\cdot |\sU|)}$:
\begin{equation}
\label{eqn:score_matrix}
    \mS_{an} = \left[ \va_1 u_1, \va_1 u_2, \ldots, \va_{|\sL|} u_{|\sU|} \right] = [ \ldots, \va_i u_j, \ldots ] 
\end{equation}
for all \mbox{$i \in \{1,2,\ldots,|\sL|\}$} and \mbox{$j \in \{1,2,\ldots ,|\sU|\}$}. To estimate the dot-product with a query, $\vq$, we can compute
\begin{equation}
\label{eqn:precompute_qk}
    \vs_{qan} = \mS^{\top}_{an}\vq
\end{equation}
where $\vs_{qan} \in \sR^{|\sL| \cdot |\sU|}$. 
Then, we can select the scores based on the $|\sL|$ indices for a given biasing phrase, and sum them to get the desired dot-product between $\vq$ and $\mW_k\vz$, which approximates $\vq^{\top}\vk$.

In sum, we use Eq.~\eqref{eqn:precompute_qk} to compute all possible dot-products between queries (acoustic encoder frames) and the keys (quantized contextual embeddings after FSQ). Then, we perform index selection and sum reduction to approximate $\mQ\mK^{\top}$ in Eq.~\eqref{eqn:cross_attention}. Finally, for each time frame we can retrieve the phrases with the highest dot-product scores. Algorithm~\ref{alg:fsq_cross_attention} describes this process in pseudo-code, where $\mE$ denotes the matrix that stores all the integer FSQ values corresponding to all the biasing phrases 
(the results of Eq.~\eqref{eqn:fsq_bound_round} for \mbox{$G=1$}). 
Note that, once we have pre-calculated the score matrix $\mS_{an}$, in practice we can run the algorithm in batches, with respect to both the time dimension (\texttt{for t = 1 to T}) and the biasing list (\texttt{for j = 1 to |B|}).

\begin{algorithm}
\caption{FSQ-based dot-product computation and retrieval}\label{alg:fsq_cross_attention}
\begin{algorithmic}
\Require $\mA \leftarrow \mW_k \mA_{\text{out}}$
\Require $\sL \leftarrow$ FSQ level information
\Require $\sU \leftarrow$ all possible values in FSQ codebooks
\Require $\mS_{an} \leftarrow [\va_1 u_1, \ldots, \va_{|\sL|} u_{|\sU|}]$
\Function{TopK}{$\mX$: Mat[$T$,$D$], 
                   $\mE$: Mat[$|\sB|$,$|\sL|$], 
                   $K$: int}
    \State $\mQ \leftarrow \mX \mW_q$
    \State $\mR \leftarrow$ Mat[$T$,$K$]
    \For{$t=1$ to $T$}
        \State $\vs^t_{qan} \leftarrow \mQ[t,:]^{\top}\mS^{\top}_{an}$
        \State $\mC \leftarrow$ Zeros($\vert \sB \vert$)
        \For{$j=1$ to $|\sB|$}
            \For{$l=1$ to $|\sL|$}
                \State $e_{jl} \leftarrow \mE[j,l]$
                \State $\mC[j] \leftarrow \mC[j] + \vs^t_{qan}[l\cdot|\sU| + e_{jl}]$
            \EndFor
        \EndFor
        \State $\mR[t] \leftarrow$ the indices of top $K$ values from $\mC$
    \EndFor
    \State return $\mR$
\EndFunction
\end{algorithmic}
\end{algorithm}

Since we are interested in the retrieval scenario, in this section we focused on the quantization of the keys and the efficient dot-product estimation between keys and queries. Of course, since FSQ is applied on the contextual embeddings, and both keys and values are linear projections of those same embeddings, quantization of the values is also possible for a full quantized cross-attention. 

\subsection{Asymptotic runtime and space complexity}
\label{subsec:compute_memory_theory}

\noindent
Table~\ref{tbl:complexity} summarizes the time and space complexity of the direct dot-product and our proposed approaches, when the size of the biasing list is the dominant factor, i.e., \mbox{$|\sB| \gg T, |\sU|, |\sL|, D, G$}. The space complexity analysis would need to consider storage of queries, keys, and the output. The queries are the acoustic encodings and require $O(TD)$ space, which can be ignored when $|\sB|$ is dominant. The output would require $O(T|\sB|)$ space, if we wanted to store all the dot-products between the acoustic encodings and the biasing phrases. This is required in case of full cross-attention, in order to calculate the softmax. However, in case of retrieval, we can calculate and store dot-products in batches and only keep the TopK at every step, so we do not consider this for our analysis. 

\begin{table}[tbh]
\let\TPToverlap=\TPTrlap
\renewcommand{\arraystretch}{1.1}
\centering
\caption{Time \& space complexity for dot-product computation between queries and keys with direct and proposed approaches.}
\label{tbl:complexity}
\begin{threeparttable}
\begin{tabular}{lcc} 
Algorithm & Runtime & Space \\ \toprule 
Direct dot-product & $O(T|\sB|D)$ & $O(|\sB|D)$ \\ 
Quantized dot-product & $O(T|\sB||\sL|G)$ & $O(|\sB||\sL|G)$ \\ 
\bottomrule
\end{tabular}
\begin{tablenotes}
    \item The size of biasing list, $|\sB|$, is assumed as dominant factor.
\end{tablenotes}
\end{threeparttable}
\end{table}

Our proposed algorithm reduces memory consumption for the keys because we no longer store a \mbox{$D$-dimensional} vector for every biasing phrase. Instead, each phrase is represented by \mbox{$|\sL|\cdot G$} indices, so the space to store the keys becomes $O(|\sB||\sL|G)$. We use these indices to retrieve the scores from the matrix $\mS_{an}$ in Eq.~\eqref{eqn:score_matrix}, which requires $O(|\sL||\sU|D)$ space, and can be ignored when $|\sB|$ is dominant.\footnote{Note that the total space needed for the score matrix is not affected by the number of groups, $G$, since each group operates on a sub-vector in $\sR^{(D/G)}$.} As a result, our proposed algorithm could save memory when \mbox{$D > |\sL|\cdot G$}.

In practice, memory savings will be even more pronounced since we only need \mbox{2 -- 3} bits to represent each of the $|\sL|$ elements in the integer FSQ representation \mbox{$\ve^g\;\forall g$} (Eq.~\eqref{eqn:fsq_bound_round}). Therefore, it is possible to represent a biasing phrase with $G$ \mbox{16-bit} integers, i.e., $G \times 2$ bytes. However, for the vanilla dot-product approach, we would still need $D$ floating point numbers for each biasing phrase, which is significantly larger even with low precision representation. This advantage of our proposed algorithm comes from the indexing structure that allows for a more compact representation.

To estimate the runtime complexity of our proposed algorithm, note that the score matrix $\mS_{an}$ can be pre-computed before inference starts. Therefore, we only need to consider the dot-product computation between the queries and $\mS_{an}$, which takes $O(T|\sL||\sU|D)$ time. Then, the index selection part would take $O(T |\sB| |\sL| G )$ time (see Algorithm~\ref{alg:fsq_cross_attention} for \mbox{$G=1$}), which is the dominant factor when $|\sB|$ is dominant. Therefore, the runtime complexity of our proposed algorithm would be lower when \mbox{$D > |\sL|\cdot G$}, similarly to space complexity. 
\section{Experiments}
\label{sect:expt}

\noindent
We carry out the experiments on a large-scale dataset consisting of examples from two tasks; dictation and assistant. Following~\cite{var_masking}, the model parameters are first estimated on semi-supervised data for a total of 500k updates, and then the contextual model is fine-tuned for another 100k updates on supervised data. In both stages, gradients are accumulated over 6,144 examples. We use SyncSGD + Adam~\cite{kingma2014adam} for distributed optimization, with exponentially decaying learning rates. The semi-supervised portion of the data consists of around 600,000 hours of automatically transcribed audio, while the supervised portion comprises about 50,000 hours of human-graded English queries; all data are anonymized at the user level. For the quantized models, we freeze all the parameters and we only train the quantizers for another 100k updates. When training contextual models, similarly to~\cite{bleeker2023approximate}, we sample 3 biasing phrases for each utterance (one positive and two distractors), then we share contextual phrases across all other utterances within a batch.

The backbone of our system is a CTC-AED model~\cite{KimHW17}. The acoustic encoder is a network with a Conv2D module, resulting in 6-fold downsampling of the input, followed by 12 conformer blocks. The AED decoder is a \mbox{3-block} bi-directional transformer.\footnote{In practice, this is implemented with 3 blocks of forward transformers and 3 blocks of backward transformers. While all 6 blocks are used during training, the backward transformers are disabled during inference.} Both the encoder and decoder blocks have a hidden dimension of 2,048 and employ an 8-head self-attention. 
The context encoder comprises 3 transformer blocks with a hidden dimension of 512 and an 8-head self-attention. The biasing module is based on single-head cross-attention, since initial experimentation showed no benefits by employing multi-head attention. The input audio is enhanced by spectral augmentation~\cite{park2019specaug} and is represented by \mbox{80-dimensional} logmel features, extracted every 10msec. The text representation is based on a SentencePiece tokenizer~\cite{kudo-richardson-2018-sentencepiece} with a vocabulary size of 6k, trained on the transcripts of the supervised portion of our data. For LM-based experimentation, we employ an internal transformer-based neural network language model (NNLM) using the same 6k tokenizer, trained on the same transcribed data, as well as the public LLM OpenLLaMA~3B~v2~\cite{openlm2023openllama,touvron2023llama} with a vocabulary size of 32k.

Models are evaluated using a test set containing 42 hours of data with queries targeted to a smart voice assistant. Around 50\% of the test set comprises contextual queries containing contact, app, and media names, anonymized at the user level. 
The remainder portion of test data consists of examples that are generic in nature and are not expected to benefit from auxiliary biasing information. We report ASR results using two metrics, word error rate (WER~[\%]) for quantifying an average system performance, and named entity error rate (\neer~[\%]) to better capture contextualization performance. We further distinguish between \neer on contact names---that comprise the biggest portion of the biasing catalogues in our test set---and on all other (non-contact) entities. Note that NEER is a binary metric getting a score of 0 only when ASR correctly predicts the complete entity filler and 100 otherwise. 

To showcase the flexibility of our systems with respect to the decoder applied, we perform experiments in both non auto-regressive and auto-regressive fashion. For the former, we apply CTC beam search followed by attention rescoring~\cite{yao2021wenet}, the latter relies on joint CTC-attention decoding~\cite{hori-etal-2017-joint}. Both cases use a beam size of 10 and a relative CTC weight of 0.3.

\subsection{Retrieval performance evaluation}
\label{subsec:exp_retrieval}

\noindent
In order to use the (quantized) NCB model to ground biasing catalogues on audio, we need to confirm that the attention head of the contextual cross-attention attends indeed to the expected entries. To do so, we feed to the NCB model an utterance with $T$ frames that contains the contextual phrase $\hat{p}$ in the reference text, together with the corresponding biasing list $\sB$ (with \mbox{$\hat{p}\in\sB$}). At every frame $t$ of the utterance we take the attention scores and we get the $K$ phrases from the biasing list with the TopK scores (Fig.~\ref{fig:retrieval_ncb}). Let's denote this set of phrases as \mbox{$\sP^t_K = \{p_i^t\}_{i=\{1..K\}}$}. Note that we treat the back-off token as an empty phrase, so if the Top1 for frame $t'$ is the back-off, then \mbox{$\sP^{t'}_1=\emptyset$}. We then combine all the extracted phrases, for all the frames, into a set \mbox{$\sS_K = \sP^1_K \cup \sP^2_K \cup \cdots \cup \sP^T_K$}, and we calculate the \emph{success rate} (or recall) as the rate for which \mbox{$\hat{p}\in \sS_K$} across our dataset. 

\begin{figure}[!htb]
\centering
\includegraphics[trim={5.3cm 15.8cm 5.4cm 4.4cm},clip,width=.45\textwidth]{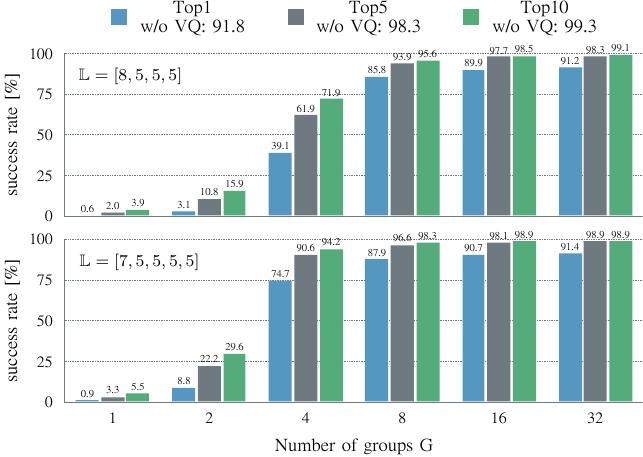}
\caption{Retrieval success rates for Top1, Top5, and Top10, for various FSQ settings. The baseline numbers (w/o quantization) are given in the legend.}
\label{fig:retrieval_success_rates}
\end{figure}

\begin{figure}[!htb]
\centering
\includegraphics[trim={5.3cm 15.8cm 5.4cm 4.4cm},clip,width=.45\textwidth]{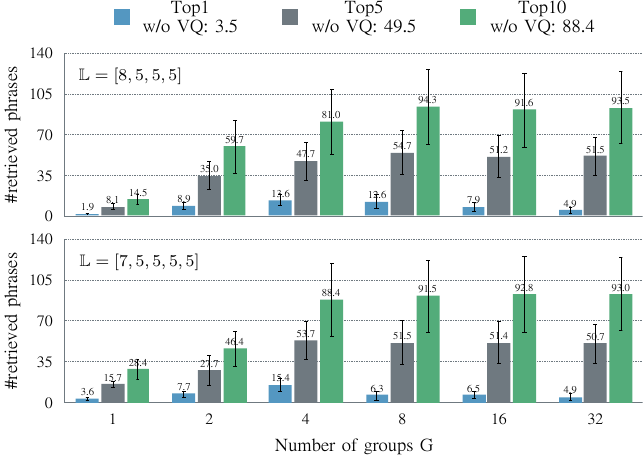}
\caption{Average number of phrases retrieved per utterance for Top1, Top5, and Top10 retrieval, for various FSQ settings. The baseline (w/o quantization) numbers are given in the legend. The error bars represent one standard deviation from the averages.}
\label{fig:retrieval_mean_phrases}
\end{figure}

We evaluate the retrieval capabilities of the models on a random subset of 1.4k examples drawn from the test set, all of which contain some contextual phrase (contact, app) anonymized at the user level, stored in the user's biasing list. For this experiment we limit the maximum length of biasing catalogues to $\max|\sB|=5$k. We report the success rates for various FSQ configurations in Fig.~\ref{fig:retrieval_success_rates}, while in Fig.~\ref{fig:retrieval_mean_phrases} we report the number of retrieved phrases per utterance (i.e., the cardinality of the set $\sS_K$ per our previous notation).

Success rates for the quantized models degrade a lot for \mbox{$G<8$}, but appear to be very close to the non-quantized baselines for \mbox{$G\geq16$}. This shows that we can effectively use the quantization-based system to retrieve a small subset of phrases without hurting ASR accuracy (see also experiments of Section~\ref{subsec:exp_asr}). For the subsequent experiments, we mainly focus on the model with \mbox{$G=16$}, \mbox{$\sL=[8,5,5,5]$}, as a good trade-off between retrieval accuracy and compute/memory gains based on Algorithm~\ref{alg:fsq_cross_attention}. The entire retrieved biasing list per utterance for this model contains on average 7.9 phrases (\mbox{max=35}) for Top1 retrieval, 51.2 phrases (\mbox{max=162}) for Top5 retrieval, and 91.6 phrases (\mbox{max=298}) for Top10 retrieval. 

\begin{figure}[!htb]
\centering
\includegraphics[trim={5.4cm 16.2cm 5.5cm 4.5cm},clip,width=.45\textwidth]{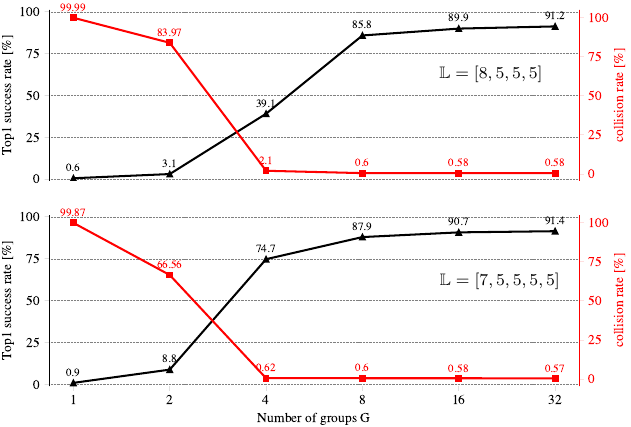}
\caption{Collision vs. Top1 retrieval success rate, for various FSQ settings.}
\label{fig:retrieval_collisions}
\end{figure}

Since we are introducing a VQ-based approach, it is important to examine potential collisions. A collision happens when different biasing phrases get the same quantization indices, meaning that the model can no longer distinguish them. Here we estimate the collision rate as 
\begin{equation}
    \text{collision} =  \frac{ \#\text{unique phrases} - \#\text{unique indices }}  {\#\text{unique phrases}}
\end{equation}
for all the phrases across all the biasing catalogues in our set. Note that in any case we include exactly $K$ phrases in the TopK retrieved list (even when multiple biasing phrases collide). As we can see in Fig.~\ref{fig:retrieval_collisions}, the collision rate is negatively correlated with the retrieval success rate, but is consistently low for \mbox{$G\geq8$}. After inspecting some collisions in those cases, most of them occur between pairs of phrases with different word ordering (e.g.,~\emph{Vector Quant} vs.~\emph{Quant Vector}), different punctuation (e.g.,~\emph{Vector Quant} vs.~\emph{Vector Quant.}), or slightly different spelling (e.g.,~\emph{Vector} vs.~\emph{Vecctor}). The very high collision rates and, hence, the low success rates for \mbox{$G=1$} are not surprising. The subset under examination contains about 700k unique biasing phrases, whereas the maximum capacity of the quantizer for $G=1$ is equal to \mbox{$\prod_{i=1}^{|\sL|}l_i$}, which is only 1,000 for \mbox{$\sL = [8,5,5,5]$} and 4,375 for \mbox{$\sL = [7,5,5,5,5]$}.

\subsection{ASR performance evaluation}
\label{subsec:exp_asr}

\noindent
We start our ASR evaluation with Table~\ref{tab:baselines}, where the first block shows baseline results for a few NCB-enabled models depicted in Fig.~\ref{fig:retrieval_ncb_vanilla}. In particular, we evaluate NCB consuming different numbers of biasing phrases, with $\max|\sB|=0$ denoting a non-biased baseline. \emph{Dense NCB} is the system that does biasing by computing full scaled dot-product cross-attention scores. When ingesting a random selection of up to 1k or 5k biasing entries,\footnote{This means that if the size of the biasing catalogue is larger than 5k (1k), we trim it to 5k (1k) randomly selected phrases.} we observe an overall 21\% relative WER reduction and between 50-55\% relative contact \neer reduction. In this work we assume $\max|\sB|=5$k as a reasonable practical upper limit on the size of the biasing inventory for edge deployment, similarly to many other works using similar techniques~\cite{chang2021context, sun2021tree, meng2024text}. However, the size of biasing catalogue may vary between users and applications in substantial ways, from a few hundred to hundreds of thousands of entries or more (e.g., when biasing to non-personal entities such as media titles), highlighting the challenge of scaling Dense NCB to uncapped biasing catalogues. 

\begin{table}[!htb]
    \renewcommand{\arraystretch}{1.1}
    \centering
    \caption{WER and \neer metrics for the baseline and NCB-enabled models, and their efficient retrieval oriented setups.
    }
    \label{tab:baselines}
    \begin{threeparttable}
    \begin{tabular}{lcccc}
        \multirow{2}{*}{System} & \multirow{2}{*}{\centering $\max|\sB|$} & \multirow{2}{*}{WER[\%]} & 
        \multicolumn{2}{p{2.4cm}}{\centering \neer [\%]}  \\ \cline{4-5}
        & & & Contact & Other \\ \toprule
        Dense NCB & 0 & 7.0 & 39.4 & 15.8   \\
        Dense NCB   & 1k & 5.6 & 19.7 & 15.4  \\ 
        Dense NCB   & 5k & 5.5 & 17.7 & 15.4   \\ \midrule
        (I) Retrieval NCB & 5k & 5.5 & 17.7 & 15.4  \\
        (II) Retrieval NCB & 5k & 5.5 & 17.7 & 15.4  \\ 
        \bottomrule
    \end{tabular}
    \begin{tablenotes}
        \item System (I) based on $G=16$, $\sL=[7,5,5,5,5]$ FSQ variant.
        \item System (II) based on $G=16$, $\sL=[8,5,5,5]$ FSQ variant.
        \item $\max|\sB|$ denotes the maximum number of biasing phrases used.
        \item Results for $\text{TopK}=5$ (in case of Retrieval NCB), w/ CTC + attention rescoring decoder.
    \end{tablenotes}
    \end{threeparttable}
\end{table}

As shown in the previous section, quantized representation offers close-to-baseline retrieval performance for $\mbox{TopK} \geq 1$, thus we propose to use quantization for large-scale retrieval, followed by dense biasing only for the acoustically grounded entries. This system is depicted in Fig.~\ref{fig:retrieval_ncb_2pass} and we will refer to this operation mode as \emph{Retrieval NCB}. The reason we do not use, here, quantized cross-attention in a standalone, single-stage manner is that, in this case, we observed a higher confusion among the phrases with high attention scores that led to degraded accuracy, when compared to full cross-attention. We further explore this issue in Appendix~\ref{appdx:quant_ncb}. Results for Retrieval NCB are reported in the second block of Table~\ref{tab:baselines}, where we can observe that the retrieval-augmented systems match the performance of Dense NCB. At the same time, this approach offers much better scaling characteristics with respect to the size of the biasing catalogues, a property that we investigate in the remainder of this section.


%
Since the proposed technique allows us to increase the size of the biasing inventory with minimal memory footprint (see also Section~\ref{subsec:compute_memory_eval}), Table~\ref{tab:enumerations} reports results for the case where we consume all the available contextual information. Note that from now on we use the Retrieval NCB system~(II) from Table~\ref{tab:baselines}. The first two rows repeat the accuracy results when the maximum allowed number of biasing entries is set to 1,000 and 5,000 random entries, respectively. We can see that although this change (from 1k to 5k entries) has small impact on the overall accuracy, it is important for named entity regions, as reflected by the contact \neer scores. This suggests that a biasing approach that puts constraints on the allowed size of the biasing catalogue is sub-optimal. Running decodes with all the available biasing information (maxing out at around 22.6k for the largest catalogues in the test set) does not seem to lead to further improvements, on average, over the trimmed configuration, likely due to the fact that few catalogues are affected by the 5k size threshold.
\begin{table}[!htb]
    \renewcommand{\arraystretch}{1.1}
    \centering
    \caption{WER and \neer metrics of Retrieval NCB for limited, full, and enumerated biasing phrase lists.
    }
    \label{tab:enumerations}
    \begin{threeparttable}
    \begin{tabular} {lcccc} 
        \multirow{2}{*}{Biasing Catalogue} & \multirow{2}{*}{$\max|\sB|$} & \multirow{2}{*}{WER[\%]} & 
        \multicolumn{2}{p{2.5cm}}{\centering \neer [\%]}  \\ \cline{4-5}
        & & & Contact & Other \\ \toprule
        Limited & 1k & 5.6 & 19.7 & 15.4  \\ 
        Limited & 5k & 5.5 & 17.7 & 15.4 \\ \midrule
        Full & 22.6k & 5.5 & 17.6 & 15.5  \\
        + Enum. All   & 43.5k & 5.5 & 16.2 & 15.8 \\ 
        + Enum. Contacts  & 42.9k & 5.4 & 15.9 & 15.7    \\ 
        \bottomrule
    \end{tabular}
    \begin{tablenotes}
        \item Results for $\text{TopK}=5$, w/ CTC + attention rescoring decoder.
    \end{tablenotes}
    \end{threeparttable}
\end{table}

We then experiment with enumerating the biasing phrases into word- and word-order-level combinations. For example, in a phone-book like scenario we assume that the user may utter arbitrary combinations of first and last names.\footnote{A single \textit{Joe Foe} phrase becomes a set \{\textit{Joe Foe}, \textit{Joe}, \textit{Foe}, \textit{Foe Joe}\}.} Such an enumeration strategy doubles the maximum inventory to around 43,000 biasing entries, but reduces \neer on contacts by 9.7\% relative when compared to the non-enumerated variant. 
Our efficient retrieval algorithm allows us to consume fully enumerated catalogues easily, without the need to cap them due to algorithmic complexity. 
Applying enumerations on all the biasing entries did not bring further gains, which was expected, since for entities such as aggregated song titles, app names, etc., enumerations are less important.


Table~\ref{tab:topk} shows 
results for different TopK retrieval settings. We observe that most of the gain is realized for $K=1$, and contact \neer saturates after $K=5$. The latter will be our default setting in the remainder of this work.

\begin{table}[!htb]
    \renewcommand{\arraystretch}{1.1}
    \centering
    \caption{WER and \neer metrics of Retrieval NCB for various TopK settings.}
    \label{tab:topk}
    \begin{threeparttable}
    \begin{tabular}{cccc}
        \multirow{2}{*}{\hspace*{.1cm}TopK\hspace*{.1cm}} & \multirow{2}{*}{WER[\%]\hspace*{.1cm}} & 
        \multicolumn{2}{p{2.5cm}}{\centering \neer [\%]}  \\ \cline{3-4}
        & & Contact & Other \\ \toprule
        0 & 7.0 & 39.4 & 15.8  \\
         1  & 5.5 & 18.3 & 15.3 \\ 
         2  & 5.5 & 17.8 & 15.4    \\ 
         5  & 5.5  & 17.6 & 15.5 \\ 
         10 & 5.5 & 17.6 & 15.4 \\ 
         20 & 5.5 & 17.6 & 15.4  \\
         \bottomrule
    \end{tabular}
    \begin{tablenotes}
        \item Using the \emph{Full biasing info} system from Table~\ref{tab:enumerations}.
        \item TopK $=0$ essentially means that NCB is disabled.
    \end{tablenotes}
    \end{threeparttable}
\end{table}

Our proposed Retrieval NCB approach is agnostic to the exact decoding mechanism applied. While results in the previous tables were based on CTC decoding with attention rescoring, 
the \emph{Retrieval NCB} rows of Table~\ref{tab:llm} report results with auto-regressive joint CTC-attention decoding~\cite{hori-etal-2017-joint}.\footnote{We still use the same streaming conformer acoustic encoder, but the attention module has access to the global set of audio encodings.} We can see that Retrieval NCB works well in this configuration as well, offering an additional 16.4\% relative reduction in contact \neer, or 14.8\% relative reduction in general WER, when compared to the attention rescoring setup (Table~\ref{tab:enumerations}).
%
%
\begin{table}[!htb]
    \renewcommand{\arraystretch}{1.1}
    \centering
    \caption{WER and \neer metrics for systems with LLM fusion and prompting after joint CTC-attention decoding.}
    \label{tab:llm}
    \begin{threeparttable}
    \begin{tabular}{lcccc}
        \multirow{2}{*}{System} & \multirow{2}{*}{$\max|\sB|$} & \multirow{2}{*}{WER[\%]} & 
        \multicolumn{2}{p{2.5cm}}{\centering \neer [\%]}  \\ \cline{4-5}
        & & & Contact & Other \\ \toprule
        Retrieval NCB     & 0 & 6.3 & 38.2 & 12.9   \\ 
        + SF NNLM        & 0 & 6.4 & 38.2 & 12.6 \\
        + DF OpenLLaMAv2 & 0 & 6.4 & 37.8 & 11.3   \\ 
        ++ prompt & \hspace*{.15cm}42.9k$^*$ & 5.9 & 31.8 & 11.0   \\ \midrule 
        Retrieval NCB     & 42.9k & 4.6 & 13.3 & 12.5   \\ 
        + SF NNLM        & 42.9k & 4.7 & 13.6 & 12.1 \\
        + DF OpenLLaMAv2 & 42.9k & 4.7 & 13.3 & 11.1   \\
        ++ prompt   & 42.9k & 4.5 & 11.0  & 11.1  \\ 
        \bottomrule
    \end{tabular}
    \begin{tablenotes}
        \item $^*$used retrieved entries only via LLM prompting path for biasing.
        \item Using the \emph{Enum. Contacts} system from Table~\ref{tab:enumerations}.
    \end{tablenotes}
    \end{threeparttable}
\end{table}

Till now, we have considered retrieval and biasing using different configurations of the cross-attention module, as in Fig.~\ref{fig:retrieval_ncb_vanilla} and \ref{fig:retrieval_ncb_2pass}; however, shortlisted phrases can be used to contextualize the model with an arbitrary biasing machinery. One recent example relies on LLMs equipped with prompting functionality, with the biasing phrases given as a prompt to a pre-trained model~\cite{Wang2023_slm,gong2024contextual}. 
For our experimentation we pack the retrieved phrases into a prompt\footnote{The prompt was: \textit{Here is a comma separated list of acoustically grounded entities you may use when predicting next relevant word: ... }} and use the LLM in a delayed fusion mode~\cite{hori2025_df} with the hypotheses emitted by CTC-attention decoders (Fig.~\ref{fig:retrieval_ncb_prompt}). Delayed fusion is a variant of shallow fusion that computes and applies LM scores for partial ASR hypotheses, but after pruning and re-tokenization of the hypotheses. This reduces the number of LLM inference calls and allows us to use an arbitrary LLM, potentially with a tokenization different than the one employed by our main ASR system. 
This means we do not need to re-train our ASR system to match the---typically---much larger LLM vocabulary, something that not only reduces compute cost, but also maintains the robustness of the ASR model~\cite{higuchi-etal-2022-bert,hori2025_df}.

The \emph{DF OpenLLaMAv2} rows of Table~\ref{tab:llm} show results with LLM delayed fusion, but without prompting. Even though there is no reduction in general WER, we can see an improvement in entity recognition. Using a robust LM especially helps with non-personal entity regions, as expected, with up to 12.4\% relative \neer-other reduction. For completeness, we also report results for decodes with shallow token-level fusion with a relatively small in-domain LM, referred to as \emph{SF NNLM}. This model is not able to be conditioned on the prompt, but helps to quantify the overall impact of LLM fusion in generic (non-prompted) mode.

The final row of the first block of the table shows the performance of the retrieval setup, when applied to shortlist biasing phrases that are packed to prompt the LLM (Fig.~\ref{fig:retrieval_ncb_prompt}). We observe that when the LLM is kept frozen and independent from the AM, the prompting performance remains somewhat limited with respect to contact \neer, with the relative improvement over baseline only being 15.8\% (for comparison, Retrieval NCB got over 65\% relative contact \neer reduction).  Note that here we use the pre-trained weights of the LLM, without further adaptation to our use case nor allowing the LLM to access acoustic embeddings via adapter mechanism. This can perhaps explain somewhat limited gains, but LLM adaptation is out of the scope of this work.\footnote{We also experimented with the pre-trained \textit{Mistral-7B-v0.1} and instruction-finetuned \textit{Mistral-7B-Instruct-v0.1} LLM variants~\cite{jiang2023mistral7b}. These did not offer new insights; in particular we found both variants to have similar ability to follow prompt contextualization instructions on our task.}

Finally, the last row of the second block of the table shows a combined variant of retrieval-based NCB where the shortlisted biasing phrases are used both as an input to a dense cross-attention module and as an LLM prompt (Fig.~\ref{fig:retrieval_ncb_2passANDprompt}). This gives us further 17.3\% relative contact \neer reduction when compared to the strongest Retrieval NCB result, or about 71\% relative when compared to the non-biased model (w/ DF OpenLLaMAv2 and \mbox{$\max|\sB|=0$}). 
It is interesting to observe a greater contextualization ability of LLM prompting when combined with Retrieval NCB than in a standalone LLM prompting manner (relative contact \neer reduction is 17.3\% and 15.8\%, respectively). This shows that for LLM fusion / re-scoring like approaches it is crucial to have high-quality biased candidate hypotheses from AM that are more amendable to LLM contextual rewrite. This also demonstrates that LLM can be leveraged effectively without acoustic encodings, and thus may potentially allow for effective conditional LLM fusion, where LLM is lazily queried for selected subsets of challenging traffic. We leave that direction for a future work.

\subsection{Compute performance evaluation}
\label{subsec:compute_memory_eval}

\begin{figure}[!htb]
\includegraphics[width=8.5cm]{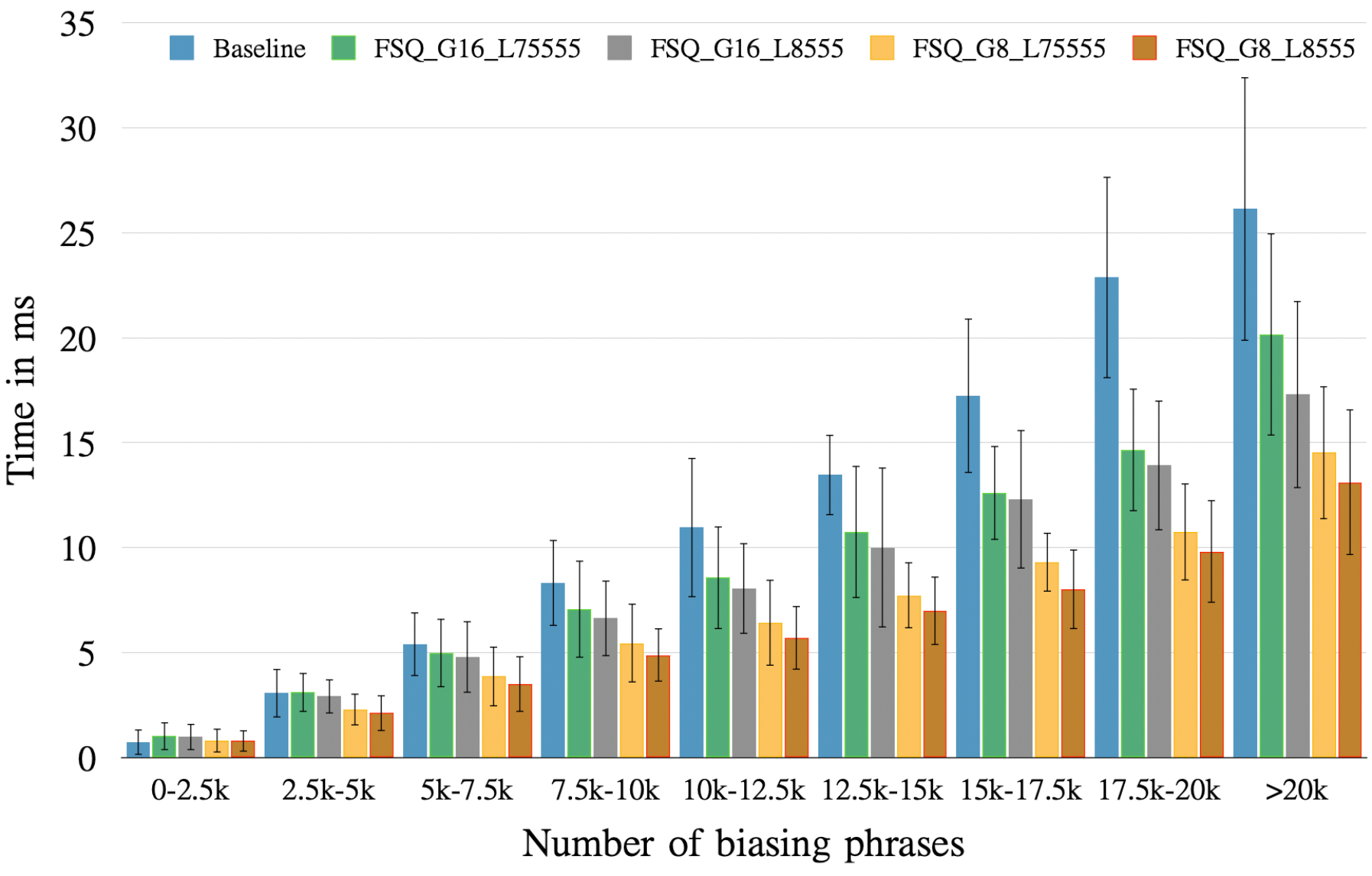}
\caption{Runtime analysis of baseline and proposed approaches across various biasing list sizes. The error bars represent one standard deviation from the averages, computed from all the queries in the same bin.}
\label{fig:vq_runtime_eval}
\end{figure}

\noindent
Fig.~\ref{fig:vq_runtime_eval} shows the runtime evaluation of the baseline and proposed approaches for dot-product estimation. 
The runtime was measured with an Intel Xeon Gold 5128 processor clocked at 2.3GHz running on a single thread. In this analysis, we measured the time for computing cross-attention up to the point of (and including) the dot-product between queries and keys during inference. Any computation that can be prepared offline, such as the linear transformation of the keys in Eq.~\eqref{eqn:qkv_linearities}, is not accounted for. For the baseline approach, this includes the query preparation and the dot-product between queries and keys in Eq.~\eqref{eqn:qkv_linearities} -- \eqref{eqn:cross_attention}. For the proposed approach, this includes the query preparation, the index selection, and the sum reduction as shown in Algorithm~\ref{alg:fsq_cross_attention}.

The results showcase that the proposed algorithm is faster than the baseline approach and the gains are bigger as the number of biasing phrases increases, which is expected since the time complexity is proportional to $|\sB|$ (Section~\ref{subsec:compute_memory_theory}). When the biasing list has over 10k entries, all FSQ configurations achieve at least 20\% reduction in runtime. For the largest biasing lists, the runtime of our proposed algorithm with the fastest FSQ setting (among the ones examined) is roughly half of the baseline. Among different FSQ configurations, smaller number of groups, $G$, and fewer levels, $|\sL|$, are contributing factors. This is consistent with the complexity analysis in Table~\ref{tbl:complexity}, where we saw that runtime complexity is proportional to both those parameters. Also note that for all the models in Fig.~\ref{fig:vq_runtime_eval}, \mbox{$|\sL|\cdot G < D = 256$}.

\begin{figure}[!htb]
\includegraphics[width=8.5cm]{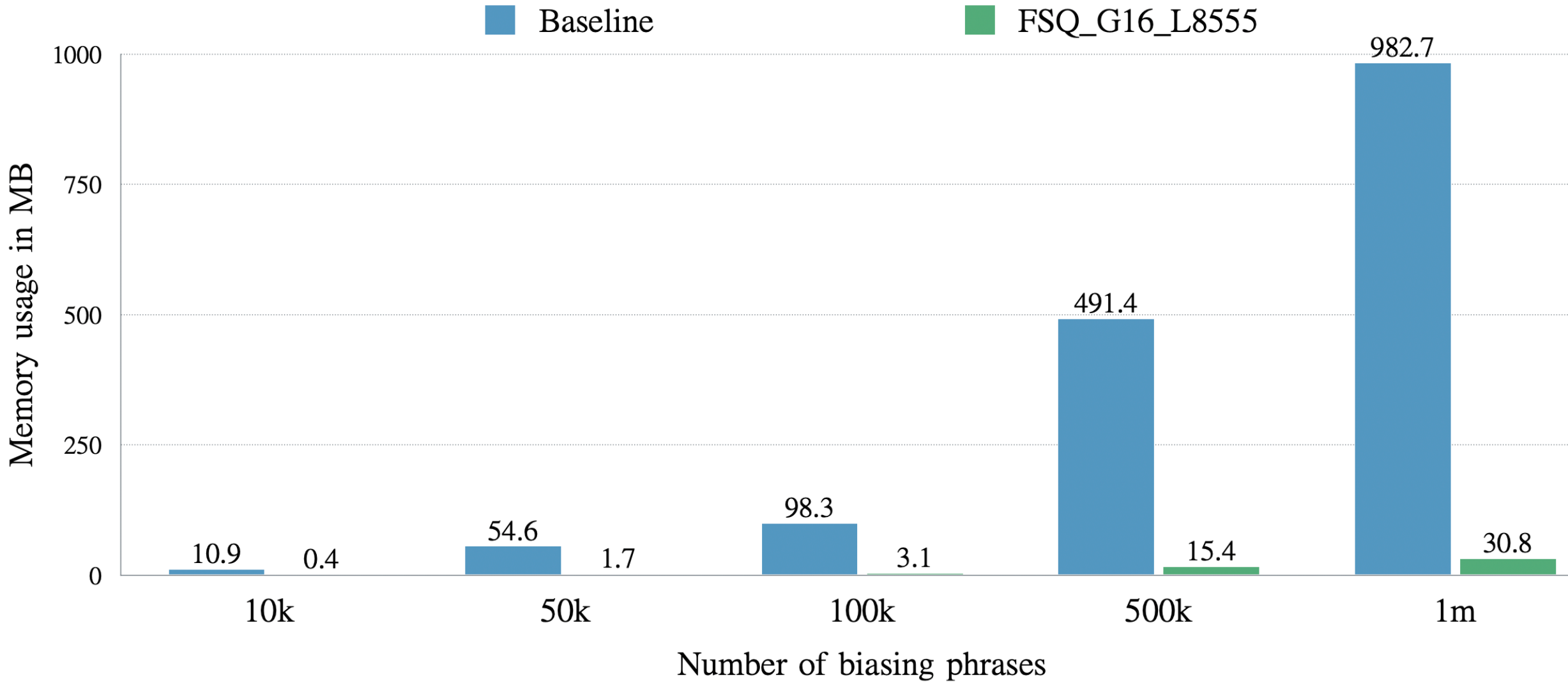}
\caption{Memory usage of baseline and proposed approaches across various biasing list sizes. The FSQ configuration has \mbox{$G=16$} groups and \mbox{$\sL = [8,5,5,5]$} levels.}
\label{fig:vq_memory_eval}
\end{figure}

Fig.~\ref{fig:vq_memory_eval} shows the memory usage analysis of the baseline and proposed approaches. For this analysis, we picked a query with about one second of audio and roughly 10k biasing phrases. Then, we artificially created larger biasing lists by repetition. The purpose of doing so is to compare memory usage in extreme hypothetical scenarios with huge biasing lists. Similar to the runtime analysis,  we measured the memory usage for computing cross-attention up to the point of dot-product between queries and keys. In this analysis, we used full precision for floating point numbers. For indices, since we only need \mbox{3 bits} for each level (to represent up to \mbox{$\max l_i=8$} values), we can use a 16-bit integer to represent all the elements $e_i^g$ in $\ve^g$, for every group \mbox{$g\in\{1,2,\ldots,G\}$} (Eq.~\eqref{eqn:fsq_bound_round}). Given that this particular FSQ configuration has 16 groups, each biasing phrase only needs 32 bytes, which is substantially smaller than the space needed for the baseline approach.

The results show that our proposed algorithm uses little memory compared to the baseline, as expected. Consider the case with 1 million biasing phrases: we need $16 \times 2 \times 1$ million bytes, i.e., 30.5MB of memory, to represent all the biasing phrases. In comparison, the baseline approach would need $256 \times 4 \times 1$ million bytes, i.e. 976.6MB of memory to represent the whole biasing list. Even if we reduced the precision for the biasing phrases from 32-bit to 8-bit, the memory needed for the baseline approach would still be much larger than our quantization-based algorithm (244MB vs. 31MB). Therefore, depending on the precision, our approach achieves over 85-95\% reduction in memory usage when the biasing list has over 10k entries. This memory reduction is due to a much more compact representation of the biasing phrases, as explained in Section~\ref{subsec:compute_memory_theory}. However, we would like to mention that to achieve the full memory reduction, one would need to implement a kernel that combines index selection, sum reduction and TopK retrieval as in Algorithm~\ref{alg:fsq_cross_attention}. Without this kernel, the memory reduction is smaller; we discuss this in detail in Appendix~\ref{appdx:implementation}. 

\section{Conclusions}
\label{sect:conclusions}

\noindent
We proposed an efficient approximation to cross-attention that uses vector quantization techniques, allowing us to ground large biasing catalogues quickly on audio data with limited memory footprint. We incorporated this mechanism in a retrieval setup, where the shortlisted phrases are used to bias an ASR system through dense cross-attention and/or through LLM prompting. We evaluated our proposed methods on a large set, with our strongest system yielding a relative contact NEER reduction of about 71\% and a relative WER reduction of about 30\%, when compared to a non-biased baseline. 

Our quantization-based technique allows the ASR model to leverage more biasing entries that would otherwise be discarded due to excessive compute and memory cost. We demonstrated the need for employing larger biasing catalogues by reporting sub-optimal results when we limited the initial catalogues to a few thousand phrases, a common practice in the literature. Our retrieval approach opens up the way to scale NCB up to a much larger number of biasing phrases, something essential in several scenarios; for example, when the goal is to bias the ASR model on the media domain (including all the aggregated song and album titles). 

While we analyzed and evaluated our system using FSQ, the ideas we described can potentially be paired with alternative efficient vector quantization algorithms and retrieval-oriented losses (e.g.,~\cite{guo2020accelerating}). Additionally, we believe that further improvements can be obtained by relatively simple enhancements. For instance, for all the LLM-based results reported in this work we used a publicly available pre-trained LLM. The overall accuracy could be improved after fine-tuning such an LLM in the speech domain and adapting it for our use case. Moreover, even though we followed the common practice of employing a separate transformer-based context encoder, we could exploit the representational capabilities of the LLM even more, by processing the biasing phrases through it in an LLM2Vec-like manner~\cite{behnamghader2024llm2vec}. This could potentially further reduce the overall memory footprint, by discarding all the additional context encoder parameters.

\section*{Acknowledgments}
\label{sect:acknowledgements}

\noindent
We would like to thank Stefan Braun, Erik McDermott, Xinwei Li and Stephen Pulman for their feedback on this work. Thanks to Xiao Zhang for providing an in-house NNLM.

\appendices
\section{Single-Stage Quantized NCB}
\label{appdx:quant_ncb}

\noindent
Throughout this paper we have explored two-stage NCB systems where we used our quantization-based approach to estimate dot-products between queries (acoustic embeddings) and keys (biasing embeddings) in order to retrieve the TopK biasing phrases with the highest scores. In this appendix we study a single-stage NCB approach where the biasing module of Dense NCB (Fig.~\ref{fig:retrieval_ncb_vanilla}) is fully replaced by its quantized counterpart; we refer to this NCB mode as \emph{Quantized NCB}.
The new quantized module approximates Eq.~\eqref{eqn:cross_attention} using discretized contextual encodings, introduced in Section~\ref{sect:formulation}, for both keys and values. Representing the keys, $\mK$, and the values, $\mV$, by their quantized approximations, $\tilde\mV$ and $\tilde\mK$, respectively, the biasing encodings can be simply estimated as 
\begin{equation}
    \tilde\mY = \softmax\left(\alpha\;\mQ \tilde\mK^{\top}\right)\tilde\mV
    \label{eqn:cross_attention_quant}
\end{equation}
Those biasing encodings, $\tilde\mY$, are added to the acoustic embeddings and fed into the ASR decoders. The dot-products in $\mQ \tilde\mK^{\top}$ are still estimated using the algorithm described in Section~\ref{sect:vq_algo}, but without the need for TopK retrieval. 

Table~\ref{tab:quant_vs_retrieval_ncb} repeats the results of Table~\ref{tab:baselines} for Dense NCB (Fig.~\ref{fig:retrieval_ncb_vanilla}) and Retrieval NCB (Fig.~\ref{fig:retrieval_ncb_2pass}) systems and compares their performance to their Quantized NCB equivalents. Even though Quantized NCB performs better that the non-biased baseline (with $\max|\sB|=0$), we can see that the quantization-based approximation, when used in a standalone manner, offers substantially worse biasing accuracy, compared to either dense cross-attention or retrieval-oriented two-stage setups. 

\begin{table}[!htb]
    \renewcommand{\arraystretch}{1.1}
    \centering
    \caption{WER and \neer metrics for the baseline and NCB-enabled models, their quantized approximations, and their retrieval oriented setups.
    }
    \label{tab:quant_vs_retrieval_ncb}
    \begin{threeparttable}
    \begin{tabular}{lcccc}
        \multirow{2}{*}{System} & \multirow{2}{*}{\centering $\max|\sB|$} & \multirow{2}{*}{WER[\%]} & 
        \multicolumn{2}{p{2.4cm}}{\centering \neer [\%]}  \\ \cline{4-5}
        & & & Contact & Other \\ \toprule
        Dense NCB & 0 & 7.0 & 39.4 & 15.8   \\
        Dense NCB   & 5k & 5.5 & 17.7 & 15.4   \\ \midrule
        (I) Quantized NCB & 5k & 6.2 & 26.6 & 16.0  \\ 
        (II) Quantized NCB & 5k & 6.3 & 28.6 &  15.9 \\ \midrule 
        (I) Retrieval NCB & 5k & 5.5 & 17.7 & 15.4  \\
        (II) Retrieval NCB & 5k & 5.5 & 17.7 & 15.4  \\ 
        \bottomrule
    \end{tabular}
    \begin{tablenotes}
        \item Systems (I) based on $G=16$, $\sL=[7,5,5,5,5]$ FSQ variant.
        \item Systems (II) based on $G=16$, $\sL=[8,5,5,5]$ FSQ variant.
        \item Results for $\text{TopK}=5$ (in case of Retrieval NCB), w/  CTC + attention rescoring decoder.
    \end{tablenotes}
    \end{threeparttable}
\end{table}

We observed that, on average, the attention scores of the quantized systems have lower values and a less peaky behavior compared to dense cross-attention. In other words, the attention probability mass, after the softmax function of Eq.~\eqref{eqn:cross_attention} or Eq.~\eqref{eqn:cross_attention_quant}, is distributed across more biasing phrases in the case of quantized approaches, which can lead to higher confusion among different phrases. This also explains the stats provided in 
Fig.~\ref{fig:retrieval_mean_phrases}, where we can see that the number of Top1 phrases retrieved per utterance, on average, is much higher for the quantized systems. For instance, for the setup we have mostly studied through our experiments in Section~\ref{sect:expt}, with \mbox{$G=16$}, \mbox{$\sL=[8,5,5,5]$}, we have 7.9 phrases retrieved on average per utterance, compared to only 3.5 phrases in the case of non-quantized cross-attention. Note that this behavior can be especially problematic during frames without spoken entities where the model is expected to attend to the back-off token. This is because if the model does not attend as much as it should to the back-off, it can yield over-biased transcriptions. However, as long as the right phrases are included in the retrieved set (i.e., the success rate---or recall---is high enough), we can successfully use the quantization-based, retrieval-oriented systems without hurting ASR accuracy.

\section{Implementation and Memory Consumption}
\label{appdx:implementation}

\noindent
In this appendix, we explain why a direct implementation of Algorithm~\ref{alg:fsq_cross_attention} in PyTorch could not achieve full memory reduction. This is due to the fact that PyTorch API does not allow performing index selection, sum reduction and retrieval at the same time. See a simplified code snippet of Algorithm~\ref{alg:fsq_cross_attention} implemented in PyTorch,
%
\begin{lstlisting}
# K is the number of TopK entries for retrieval
# T is the length of acoustic encoder frame sequence
# B is the number of biasing phrases
# G is the number of groups
# L is the depth of the FSQ level
# U is the number of different possible values in all FSQ levels
# e contains the indices with shape [B, G*L]
# S_qan is the score tensor with shape [T, G*L*U]
# result is the output tensor with shape [T, K]

e = e.view(-1)
a = torch.index_select(S_qan, dim=-1, index=e)
a = a.view(T, B, G*L)
s = torch.sum(a, dim=-1)
s = s.view(T, B)
result = torch.topk(s, K, dim=1)
\end{lstlisting}

\begin{figure}[!htb]
\includegraphics[width=8.5cm]{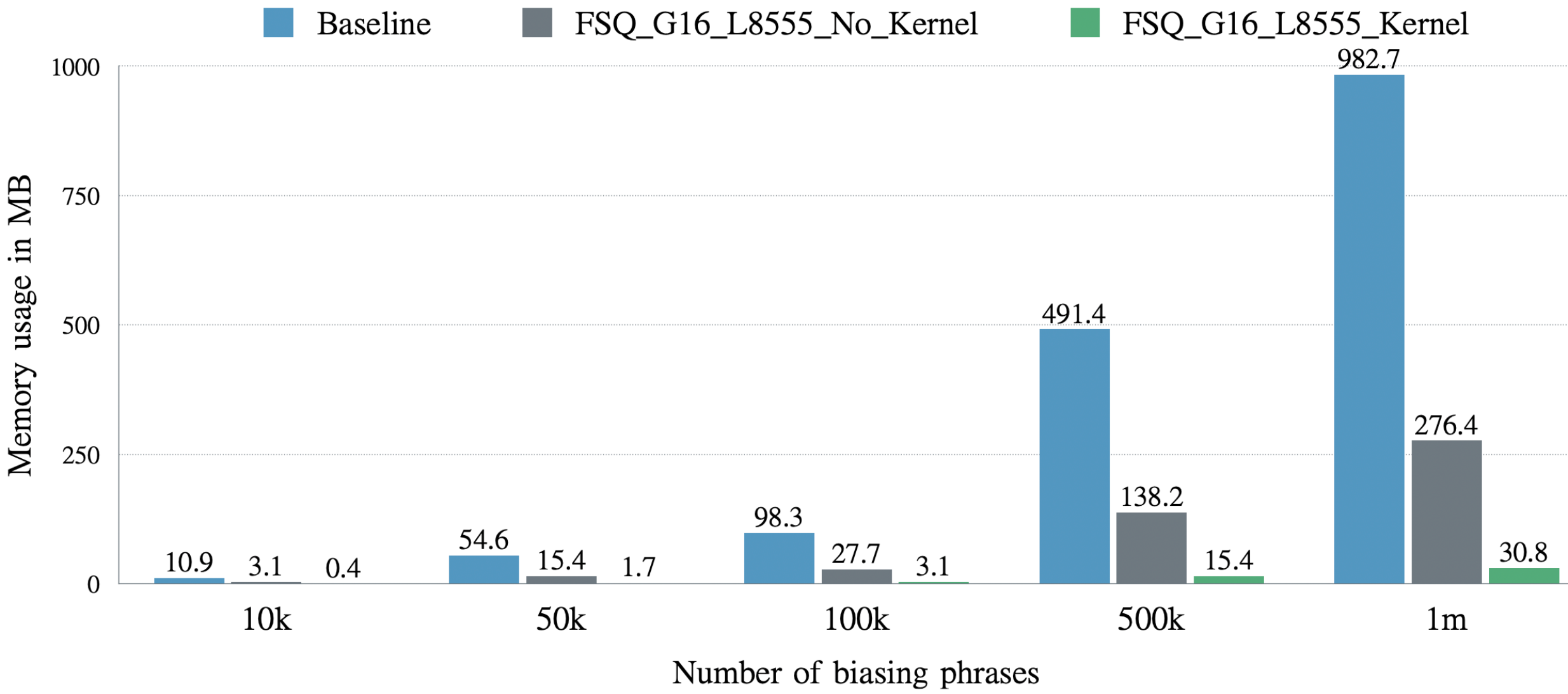}
\caption{Memory usage of baseline and our proposed approach with or without a custom kernel to perform index selection, sum reduction and retrieval at the same time. The FSQ uses \mbox{$G=16$} groups and \mbox{$\sL=[8,5,5,5]$} levels.}
\label{fig:vq_memory_eval_impl}
\end{figure}

The problem 
is that the index selection step would create an intermediate tensor, \texttt{a}, of shape \mbox{\texttt{[T, B*L*G]}}, which is dominated by \texttt{B} for large biasing catalogues. For instance, for a typical setting in our experiments we could have \texttt{B=10000}, \texttt{L=4}, \texttt{G=16}, and \texttt{T=33} for a 2sec audio (200 frames with 6-fold downsampling). Additionally, the TopK operation is performed only after we have the entire score tensor, \texttt{s}. That said, the memory consumption in that case is still smaller than the baseline approach that needs $O(|\sB| D)$ space, since \mbox{$D > G \cdot |\sL|$} in a typical setting. To achieve full memory reduction, however, one would need to implement a kernel that performs index selection, sum reduction and TopK retrieval at the same time as in Algorithm~\ref{alg:fsq_cross_attention}. Fig.~\ref{fig:vq_memory_eval_impl} compares the memory usage with or without this custom kernel,
using the same assumptions as in Section~\ref{subsec:compute_memory_eval}.

\bibliographystyle{IEEETran}
\bibliography{bib/abbrev,bib/e2e,bib/vq,bib/ml_other}

\end{document}